\documentclass[apjl]{emulateapj}

\slugcomment{accepted to ApJ Letters}

\shorttitle{DENSITY PDF: SOLENOIDAL VS COMPRESSIVE FORCING}
\shortauthors{Federrath, Klessen, \& Schmidt}

\begin{document}

\title{THE DENSITY PROBABILITY DISTRIBUTION IN COMPRESSIBLE ISOTHERMAL TURBULENCE: SOLENOIDAL VS COMPRESSIVE FORCING}
\author{Christoph Federrath\altaffilmark{1,2}, Ralf S.~Klessen\altaffilmark{1}, and Wolfram Schmidt\altaffilmark{3}}

\email{chfeder@ita.uni-heidelberg.de}
\email{rklessen@ita.uni-heidelberg.de}
\email{schmidt@astro.uni-wuerzburg.de}

\altaffiltext{1}{Zentrum f\"ur Astronomie der Universit\"at Heidelberg, \\Institut f\"ur Theoretische Astrophysik, Albert-Ueberle-Str.~2, D-69120 Heidelberg, Germany}
\altaffiltext{2}{Max-Planck-Institute for Astronomy, K\"onigstuhl 17, D-69117 Heidelberg, Germany}
\altaffiltext{3}{Lehrstuhl f\"ur Astronomie der Universit\"at W\"urzburg, \\Institut f\"ur Theoretische Physik und Astrophysik, Am Hubland, D-97074 W\"urzburg, Germany}

\begin{abstract}
The probability density function (PDF) of the gas density in turbulent supersonic flows is investigated with high-resolution numerical simulations. In a systematic study, we compare the density statistics of compressible turbulence driven by the usually adopted solenoidal forcing (divergence-free) and by compressive forcing (curl-free). Our results are in agreement with studies using solenoidal forcing. However, compressive forcing yields a significantly broader density distribution with standard deviation $\sim\!3$ times larger at the same rms Mach number. The standard deviation-Mach number relation used in analytical models of star formation is reviewed and a modification of the existing expression is proposed, which takes into account the ratio of solenoidal and compressive modes of the turbulence forcing.
\end{abstract}


\section{INTRODUCTION}
The pioneering works by \citet[][PNJ97 below]{PadoanNordlundJones1997} and \citet[][PV98 below]{PassotVazquez1998} have shown that the standard deviation (stddev) $\sigma_\rho$, i.e., the width or dispersion of the linear density PDF $p_\rho$ grows proportional to the rms Mach number $\mathcal{M}$ of the turbulent flow, however, with proportionality constants in disagreement by a factor of $2$. The exact dependence of the PDF's stddev on the rms Mach number is a key ingredient for analytical models of star formation. For instance, \citet{PadoanNordlund2002} and \citet{HennebelleChabrier2008} relate the density PDF to the core mass function (CMF) and stellar initial mass function (IMF). \citet{Tassis2007} uses the density PDF on galactic scales to reproduce the Kennicutt-Schmidt relation. \citet{Elmegreen2008} suggests that the star formation efficiency is a function of the density PDF.

In the present work, we solve the discrepancy between PNJ97 and PV98 by showing that the stddev of the PDF is not only a function of the rms Mach number, but is also very sensitive to the relative importance of solenoidal (divergence-free) and compressive (curl-free) modes of the turbulence forcing, leading to variations in the stddev up to factors of $\sim\!3$ for the same rms Mach number. The main result of the present work is that the conclusions of PNJ97 and PV98 can be harmonized, if we take into account that PNJ97 have analyzed purely solenoidal forcing, whereas PV98 have studied purely compressive forcing. This apparent difference has not been considered analytically or numerically before.

We begin by explaining our numerical method in \S\ref{sec:methods}. \S\ref{sec:results} shows that our results are consistent with previous studies using solenoidal forcing, whereas compressive forcing yields a PDF with stddev $\sim\!3$ times larger in 3-dimensional turbulent flows. We present a heuristic model explaining this difference, which is based on the ratio of solenoidal to compressive modes of the forcing to estimate the proportionality constant in the stddev-Mach relation. \S\ref{sec:conclusions} summarizes our conclusions.

\begin{figure*}[t]
\centerline{
\includegraphics[width=0.49\linewidth]{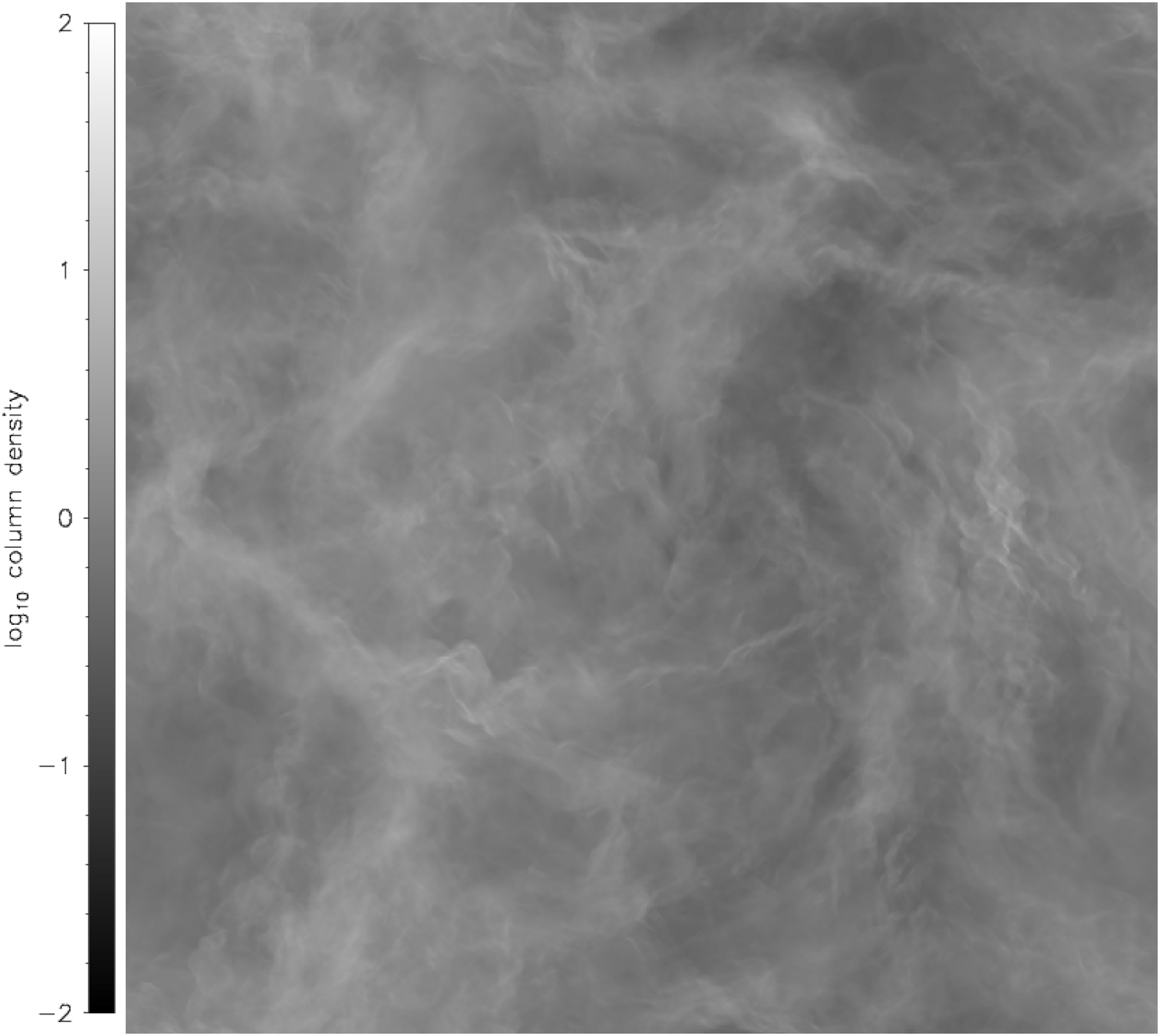}
\includegraphics[width=0.49\linewidth]{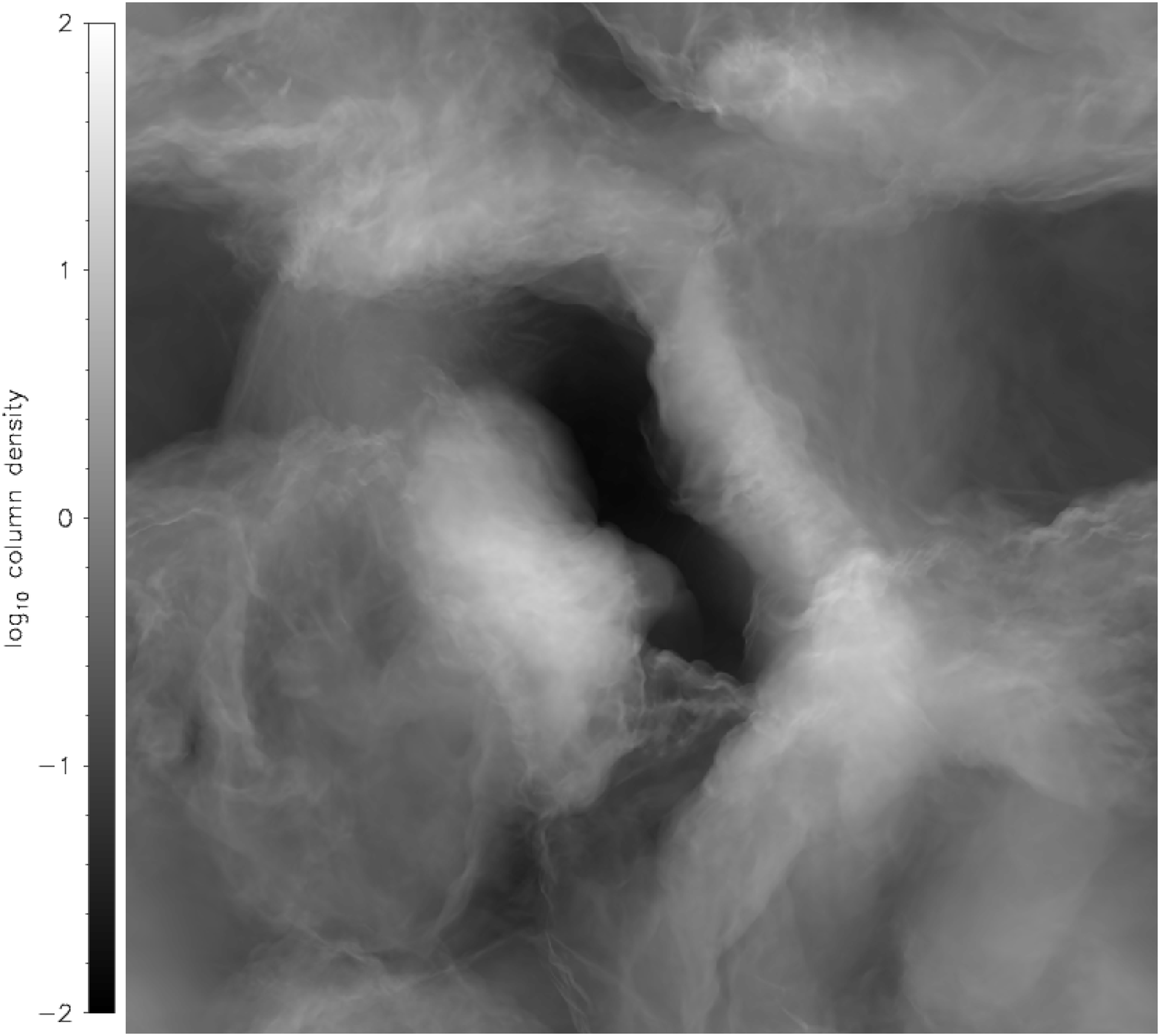}
}
\caption{Column density field in units of the mean column density for solenoidal forcing (\emph{left}), and compressive forcing (\emph{right}) at a randomly picked time in the regime of statistically stationary turbulence. Both maps show 4 orders of magnitude in column density with the same scaling for direct comparison of solenoidal and compressive forcing at rms Mach number $\sim\!5.5$.}
\label{fig:snapshots}
\end{figure*}

\section{NUMERICAL METHODS} \label{sec:methods}
The piecewise parabolic method \citep{ColellaWoodward1984} implementation of FLASH3 \citep{FryxellEtAl2000,DubeyEtAl2008} was used to solve the hydrodynamic equations on periodic uniform grids with $16384^1$, $4096^2$, and $1024^3$ grid points in 1, 2 and 3 dimension(s) (1D, 2D, 3D). Since we model isothermal gas, it is convenient to define $s\equiv\ln(\rho/\rho_0)$ as the natural logarithm of the density divided by the mean density $\rho_0$. Isothermal gas has been modelled several times in the context of molecular cloud dynamics taking periodic boundaries and studying compressible turbulence with solenoidal or weakly compressive forcing \citep[e.g.,][]{StoneOstrikerGammie1998,MacLowEtAl1998,MacLow1999,KlessenHeitschMacLow2000,HeitschMacLowKlessen2001,BoldyrevNordlundPadoan2002,LiKlessenMacLow2003,PadoanJimenezNordlundBoldyrev2004,JappsenEtAl2005,BallesterosEtAl2006,KritsukEtAl2007,DibEtAl2008,OffnerKleinMcKee2008}. Most of these studies have purely solenoidal forcing motivated by incompressible turbulence studies, some have weakly compressive forcing. The latter is a result of the natural ratio of solenoidal to compressible modes of a Gaussian field prepared in Fourier space without subsequent projection. This natural ratio is 2:1 in 3D and 1:1 in 2D. Accordingly, the ratio of compressive to the sum of compressive plus solenoidal modes is 1:3 in 3D and 1:2 in 2D. Only PV98 have purely compressive forcing, because they analyze 1D simulations where no solenoidal component exists and the ratio of compressive to total modes is 1:1.

The forcing is typically either modelled as a static pattern following the recipes by \citet{MacLowEtAl1998} and \citet{StoneOstrikerGammie1998}, or by an Ornstein-Uhlenbeck (OU) process with finite autocorrelation timescale $T$ \citep[e.g.,][]{EswaranPope1988,SchmidtEtAl2006}. Note that driving with a static pattern is the limiting case of an OU process with infinite autocorrelation timescale. We use the OU process and follow the usual approach, setting $T$ equal to the dynamical timescale $T=L/(2V)$ ($L$ is the size of the computational domain, $V=c_s\mathcal{M}$ and $\mathcal{M}$ is the rms Mach number), which is equal to the decay time constant of the turbulence \citep{StoneOstrikerGammie1998,MacLow1999} at the scales of energy injection $1<k<3$. This guarantees a well-defined stochastic driving field, that varies smoothly in space and time. We checked that modelling of the forcing as an almost static pattern (increasing $T$ by one order of magnitude) did not significantly affect our results.

In order to obtain a purely solenoidal, or a purely compressive forcing (or any combination), a Helmholtz decomposition can be made by applying the projection operator in Fourier space ($k$-space)
\begin{equation} \label{eq:projectionoperator}
\mathcal{P}_{ij}^\zeta=\zeta\mathcal{P}_{ij}^\perp+(1-\zeta)\mathcal{P}_{ij}^\parallel=\zeta\delta_{ij}+(1-2\zeta)\frac{k_i k_j}{|k|^2}\;,
\end{equation}
to the random vector field returned by the OU process or generated by the usual recipes. The parameter $\zeta\in[0,1]$ controlls the relative importance of solenoidal and compressive modes. If we set $\zeta=1$, $\mathcal{P}_{ij}^\zeta$ projects only the solenoidal component, whereas only the compressive component is obtained by setting $\zeta=0$.

\begin{figure}[t]
\centerline{\includegraphics[width=1.0\linewidth]{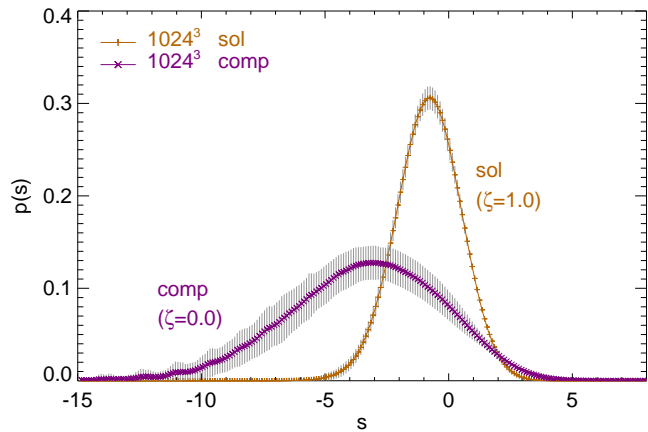}}
\caption{Volume-weighted density PDFs $p(s)$ in linear scaling where $s=\ln(\rho/\rho_0)$. The PDF obtained by compressive forcing (comp, $\zeta=0.0$) is much broader compared to the solenoidal one (sol, $\zeta=1.0$) at the same rms Mach number. The peak is shifted due to mass conservation \citep{Vazquez1994}. Gray error bars indicate 1-sigma temporal fluctuations of the PDF. A sample of $\sim\!10^{11}$ datapoints contribute to each PDF.}
\label{fig:solcompPDF}
\end{figure}

\section{RESULTS AND DISCUSSION OF THE STANDARD DEVIATION - MACH NUMBER RELATION} \label{sec:results}
Figure~\ref{fig:snapshots} compares column density maps for solenoidal vs.~compressive forcing of our 3D models ($1024^3$) from a randomly picked snapshot in the regime of statistically stationary turbulence. Obviously, compressive forcing yields much larger density contrasts, despite the fact that the rms Mach number $\mathcal{M}\!\sim\!5.5$ is about the same in both cases. This is quantitatively shown in Fig.~\ref{fig:solcompPDF}, which presents the comparison of the time averaged volume-weighted density PDFs in the solenoidally and compressively driven cases. Most importantly, although the rms Mach numbers are almost the same, the stddev $\sigma_\rho$ of the PDF obtained in compressive forcing is $\sim\!3$ times larger than for solenoidal forcing (Tab.~\ref{tab:parameters}). This is the main result of the present study and has important consequences: The stddev of the density PDF is not only a function of the rms Mach number as found by PNJ97 and PV98, but also a function of the relative strength of solenoidal to compressive modes of the turbulence forcing, i.e., a function of the projection parameter $\zeta$ in eq.~(\ref{eq:projectionoperator}). In the following, we review the stddev-Mach number relation and present a heuristic model for the proportionality constant in this relation.

\begin{table}
\begin{center}
\caption{Parameters in solenoidal and compressive forcing}
\def\arraystretch{1.2} \label{tab:parameters}
\begin{tabular}{rrlrrr}
\hline
\hline
\multicolumn{1}{c}{$D$} & \multicolumn{1}{c}{resolution} & \multicolumn{1}{c}{$\zeta$} & \multicolumn{1}{c}{$\mathcal{M}$} & \multicolumn{1}{c}{$\sigma_\rho/\rho_0$} & \multicolumn{1}{c}{$b$} \\
\hline
\noalign{\smallskip}
3 &  $1024^3$ & $1.0\,$ (sol) & $5.3^{\pm0.2}$ & $1.9^{\pm0.1}$ & $0.36^{\pm0.03}$ \\
3 &  $1024^3$ & $0.0\,$ (comp)& $5.6^{\pm0.3}$ & $5.9^{\pm1.0}$ & $1.05^{\pm0.19}$ \\
2 &  $4096^2$ & $1.0\,$ (sol) & $5.6^{\pm0.5}$ & $2.4^{\pm0.5}$ & $0.43^{\pm0.10}$ \\
2 &  $4096^2$ & $0.0\,$ (comp)& $5.7^{\pm0.6}$ & $5.0^{\pm1.5}$ & $0.88^{\pm0.28}$ \\
1 & $16384^1$ & $0.0\,$ (comp)& $5.0^{\pm1.0}$ & $4.5^{\pm2.0}$ & $0.90^{\pm0.44}$ \\
\hline
\end{tabular}
% \tablenotetext{a}{another sample footnote}
% \tablecomments{We can also attach}
\end{center}
\end{table}

It has been pointed out by PNJ97 and measured by PV98 that the stddev $\sigma_\rho$ of the PDF of the linear density $p_\rho$ grows linear with the rms Mach number $\mathcal{M}$ like
\begin{equation}
\sigma_\rho/\rho_0=b\mathcal{M} \label{eq:sigrhoofmach}
\end{equation}
with the proportionality constant $b$. PV98 find $b\!\sim\!1$ in 1D simulations with $\mathcal{M}$ ranging from subsonic to supersonic flows. PNJ97 motivate and explain eq.~(\ref{eq:sigrhoofmach}) with the isothermal hydrodynamic shock jump conditions. They assume the PDF follows a lognormal analytical form
\begin{equation}
p_s(s)=\frac{1}{\sqrt{2\pi\sigma_s^2}}\,\exp\left[-\frac{(s-s_0)^2}{2\sigma_s^2}\right] \label{eq:log-normal}
\end{equation}
to get an expression for the stddev in the logarithmic density
\begin{equation}
\sigma_s=\left[\ln\big(1+b^2\mathcal{M}^2\big)\right]^{1/2}\;. \label{eq:sigsofmach}
\end{equation}
Using $p_s ds = p_\rho d\rho$, it is easy to show that for any density PDF, whether lognormal or not, $\sigma_\rho^2=\rho_0^2\int_{-\infty}^{\infty}\left[\exp(s)-1\right]^2 p_s ds$. From the assumption~(\ref{eq:log-normal}) it follows that eq.~(\ref{eq:sigsofmach}) and~(\ref{eq:sigrhoofmach}) are equivalent and that both expressions have the same $b$. This means that eq.~(\ref{eq:sigrhoofmach}) is the basic stddev-Mach number relation from which eq.~(\ref{eq:sigsofmach}) follows, if a lognormal PDF is assumed.

PNJ97 applied eq.~(\ref{eq:sigsofmach}) to magnetohydrodynamical (MHD) simulations and obtained $b\!\sim\!0.5$. Consequently, their stddev is only half as large as that found by PV98. \citet{PadoanEtAl2007} argue that eq.~(\ref{eq:sigsofmach}) may hold for MHD turbulence as well, if $\mathcal{M}$ is replaced by the Alfv\'enic Mach number, although the shock jump conditions motivating eq.~(\ref{eq:sigrhoofmach}) are different for shocks perpendicular to the magnetic field. In MHD simulations, \citet{LiEtAl2004} and \citet{LiEtAl2008} find $b\!\sim\!0.38$ and $b\!\sim\!0.41$ for ideal MHD and $b\!\sim\!0.58$ by including ambipolar diffusion (AD). The latter means that including AD broadens the PDF over ideal MHD, because of the reduced magnetic pressure compared to ideal MHD. Therefore, the parameter $b$ in eq.~(\ref{eq:sigrhoofmach}) and~(\ref{eq:sigsofmach}) is not expected to be universal, but depends on the magnetic field \citep[see also,][]{VazquezEtAl2005}. On the other hand, \citet{LemasterStone2008} find that even strong magnetic fields do not alter the stddev-Mach number relation significantly. In the following discussion, we will concentrate on purely hydrodynamical estimates showing that $b$ is much more sensitive to the way of forcing compared to the reported MHD effects. In order to show this, we use the more fundamental stddev-Mach number relation given by eq.~(\ref{eq:sigrhoofmach}), which does not rest on the additional assumption of a lognormal density PDF.

\begin{figure}[t]
\centerline{\includegraphics[width=1.0\linewidth]{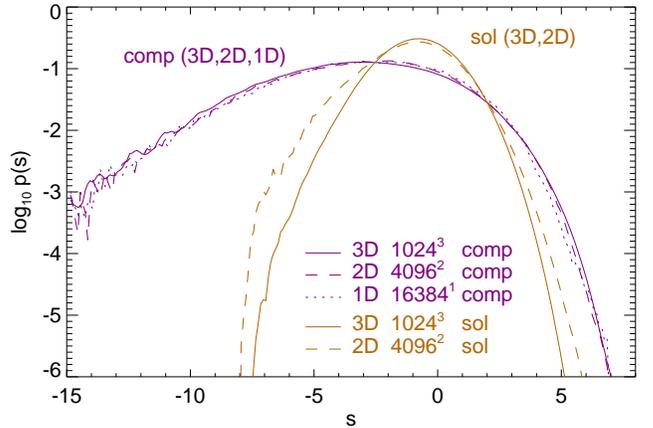}}
\caption{Density PDFs $p(s)$ obtained from 3D, 2D and 1D simulations with compressive forcing and from 3D and 2D simulations using solenoidal forcing in logarithmic scaling. Note that in 1D, only compressive forcing is possible as in the study by PV98. As suggested by eq.~(\ref{eq:b}), compressive forcing yields almost identical density PDFs in 1D, 2D and 3D with $b\!\sim\!1$ because compressive forcing induces gas compression along all available spatial directions. Solenoidal forcing on the other hand leads to a density PDF with $b\!\sim\!1/2$ in 2D and with $b\!\sim\!1/3$ in 3D as a consequence of the natural ratio of compressible to compressible plus solenoidal modes building up in the velocity field, which is 1:2 in 2D and 1:3 in 3D for solenoidally driven supersonic turbulent flows.}
\label{fig:solcompPDFdims}
\end{figure}

From purely hydrodynamic studies utilizing different numerical methods and resolutions, typically smaller but roughly consistent values of $b$ are found compared to $b\!\sim\!1$ by PV98, e.g., $b\!\sim\!0.35$ using $200,000$ SPH particles \citep{LiKlessenMacLow2003}, $b\!\sim\!0.26$ using ENZO with $1024^3$ cells \citep{KritsukEtAl2007}\footnote{In their recent $2048^3$ study with solenoidal forcing, they find a slightly larger $b\!\sim\!0.32$ (Kritsuk 2008, private communication).}, $b\!\sim\!0.37$ using another grid-based method with $512^3$ cells \citep{BeetzEtAl2008}. It is important to note that all the aforementioned studies (including the MHD studies) use solenoidal or weakly compressive forcing, except for PV98, who naturally have purely compressive forcing because they analyze 1D simulations. Similar to PV98, we find $b\!\sim\!1.05$ for purely compressive forcing (Tab.~\ref{tab:parameters}). For purely solenoidal forcing, we get $b\!\sim\!0.36$ in agreement with the estimates by \citet{LiKlessenMacLow2003}, \citet{KritsukEtAl2007} and \citet{BeetzEtAl2008}. This shows that the width of the PDF does not only depend on the rms Mach number given by eq.~(\ref{eq:sigrhoofmach}), but also on the mixture of solenoidal and compressive modes $\zeta$ of the forcing. This dependence was qualitatively mentioned by \cite{NordlundPadoan1999}, although up to now, no quantitative estimate existed.

Based on the diversity of $b$ in eq.~(\ref{eq:sigrhoofmach}) and~(\ref{eq:sigsofmach}) obtained in the studies mentioned above and based on our direct comparison of solenoidal and compressive forcing, we suggest that a plausible estimate for the proportionality constant $b$ is obtained by taking into account the compressibility induced by the forcing,
\begin{equation}
b=1+\left[\frac{1}{D}-1\right]\zeta=
\left\{
\begin{array}{lr}
1-(2/3)\zeta, & \textnormal{for $\;D=3$}\phantom{\,.}\\
1-(1/2)\zeta, & \textnormal{for $\;D=2$}\phantom{\,.}\\
1,            & \textnormal{for $\;D=1$}\,.
\end{array}
\right.
\label{eq:b}
\end{equation}
The relative strength of solenoidal to compressive modes $\zeta$ is defined in the projection operator, eq.~(\ref{eq:projectionoperator}).

For solenoidal forcing ($\zeta=1$), by construction no compression is directly induced by the forcing. Rather, density fluctuations are solely a result of compressions naturally building up in a supersonic turbulent flow. This is a consequence of the ratio of compressible modes to the sum of compressible plus solenoidal modes in the velocity field, that the turbulent gas adjusts itself to \citep[see, e.g.,][]{ElmegreenScalo2004}. This ratio is related to the number of degrees of freedom $D$ (spatial directions) available for compressible modes and is simply given by $1/D$ showing up in eq.~(\ref{eq:b}). Consequently, eq.~(\ref{eq:b}) leads to $b\!\sim\!1/3$ for the 3D case with solenoidal forcing. This is in reasonable agreement with the results by \citet{LiKlessenMacLow2003}, \citet{KritsukEtAl2007}, \citet{BeetzEtAl2008} and ours.

The 1D case as analyzed by PV98 is a special case of compressive forcing in arbitrary dimensions. Note that for compressive forcing, eq.~(\ref{eq:b}) always suggests $b\!\sim\!1$ independent of $D$. This can be understood as a consequence of the \emph{direct} compression induced by the compressive force field, contrary to solenoidal forcing for which the natural ratio of modes in the velocity field determines $b$. Compressive forcing by construction immediately induces compressions along all available spatial directions $D$ directly, and $b\!\sim\!D/D=1$. The 1D calculations ($\zeta=0$) by PV98 therefore exhibit $b\!\sim\!1$. Similar, in our high-resolution 3D case of purely compressive forcing ($\zeta=0$), compression is induced along all the $3$ available spatial directions resulting in $b\!\sim\!3/3=1$.

Besides testing eq.~(\ref{eq:b}) for the extreme cases of solenoidal and compressive forcing at different dimensionality, we also aim at testing it for intermediate values of $\zeta$ (mixtures of solenoidal and compressive forcing) using reported results in the literature and additional numerical simulations.

\citet{Vazquez1994} have analyzed 2D simulations with a 1:1 mixture of solenoidal and compressive forcing ($\zeta=0.5$). Thus, the expected value is $b\!\sim\!0.75$, and they find $b\!\sim\!0.7$ close to the expectation. \citet{SchmidtEtAl2008} use a mainly compressive forcing ($\zeta=0.1$) obtaining $b\!\sim\!0.94$ from their data, while eq.~(\ref{eq:b}) suggests $b\!\sim\!0.93$.

We ran additional 2D and 1D simulations with solenoidal and compressive forcing to support our heuristic model, eq.~(\ref{eq:b}). In order to get a statistically significant sample, the numerical resolution along the spatial directions was increased from $1024$ in 3D to $4096$ in 2D and $16384$ in 1D, as well as the integration times were increased yielding at least twice as many sampling points as PV98 have. Table~\ref{tab:parameters} summarizes the parameters of all simulations and numerical estimates for $b$. As expected from eq.~(\ref{eq:b}), the compressive cases in 1D, 2D and 3D, all exhibit nearly the same $b\!\sim\!1$. Moreover, the PDFs are very similar in all of the compressively driven cases, which is shown in Fig.~\ref{fig:solcompPDFdims}. The 3D case with solenoidal forcing is also included for comparison ($b\!\sim\!1/3$), as well as the 2D case with solenoidal forcing, for which we estimate $b\!\sim\!0.43$ close to the prediction $b\!\sim\!1/2$.

\section{SUMMARY AND CONCLUSIONS} \label{sec:conclusions}
Performing high-resolution hydrodynamical simulations of driven isothermal compressible turbulence, we have found that the way of forcing the turbulence has a strong effect on density statistics. We have compared the usually adopted solenoidal forcing (divergence-free) and compressive forcing (curl-free). The most important result is that in 3D, compressive forcing yields a density PDF with stddev $\sim\!3$ times larger compared to solenoidal forcing for the same rms Mach number.

As found by PNJ97 and PV98, the stddev $\sigma_\rho$ is increasing directly proportional to the rms Mach number, however, they find different proportionality constants. Taking into account the ratio of solenoidal to compressive modes $\zeta$ of the forcing resolves the disagreement between PNJ97 and PV98. We suggest that the proportionality constant $b$ in the stddev-Mach number relations~(\ref{eq:sigrhoofmach}) and~(\ref{eq:sigsofmach}) can be determined by a heuristic model summarized in eq.~(\ref{eq:b}). In the case of compressive forcing ($\zeta=0$), the proportionality constant $b\!\sim\!1$ irrespective of the dimensionality of the simulation. Solenoidal forcing on the other hand yields $b\!\sim\!1/3$ in 3D and $b\!\sim\!1/2$ in 2D.

We mention the impact of our results on analytical models linking the statistics of supersonic turbulence to the CMF/IMF. \citet{PadoanNordlund2002} and \citet{HennebelleChabrier2008} rely on integrals over the density PDF to get a handle on the mass of objects above a certain density threshold. Since the width of the PDF is so sensitive to the mixture of solenoidal and compressive modes $\zeta$, we also expect a strong influence on the derived CMF/IMF. Indeed, the larger dispersion obtained from compressive forcing leads to better agreement (Hennebelle 2008, private communication) of the analytic expression with the observed IMF in \citet[][]{HennebelleChabrier2008}.

Given that many proposed sources of interstellar turbulence \citep[e.g.,][]{ElmegreenScalo2004,MacLowKlessen2004} are likely to directly excite compressive modes (e.g., galactic spiral density shocks, large scale gravitational contraction, supernova explosions, protostellar jets, winds and outflows), it is reasonable to expect that turbulence in the ISM is driven by a mixture of solenoidal and compressive modes, possibly with compressive modes being more important than solenoidal modes.

\acknowledgements
We thank Mordecai-Mark Mac Low for valuable discussions on the present study. CF acknowledges financial support by the International Max Planck Research School for Astronomy and Cosmic Physics (IMPRS-A) and the Heidelberg Graduate School of Fundamental Physics (HGSFP). The HGSFP is funded by the Excellence Initiative of the German Research Foundation DFG GSC 129/1. RSK thanks for support from the Emmy Noether grant KL 1358/1. CF and RSK acknowledge subsidies from the DFG SFB 439 Galaxies in the Early Universe. The simulations used resources from HLRBII project h0972 at Leibniz Rechenzentrum Garching. The software used in this work was in part developed by the DOE-supported ASC/Alliance Center for Astrophysical Thermonuclear Flashes at the University of Chicago.

\bibliographystyle{apj}

\end{document}